\documentclass[12pt,preprint]{aastex}
\usepackage{mathptmx}  % use mathptmx to replace the previous "times"

\newcommand{\kms}{km~s$^{-1}$} %

\shorttitle{Fast-Moving Waves in Sunspots} 
\shortauthors{Zhao et al.}

\begin{document}
\title{Detection of Fast-Moving Waves Propagating Outward along Sunspots' 
Radial Direction in the Photosphere}
\author{Junwei Zhao\altaffilmark{1}, Ruizhu Chen\altaffilmark{2,1},
	Thomas Hartlep\altaffilmark{3}, Alexander G. Kosovichev\altaffilmark{4}}
\altaffiltext{1}{W.~W.~Hansen Experimental Physics Laboratory, Stanford
University, Stanford, CA 94305-4085, USA}
\altaffiltext{2}{Department of Physics, Stanford University, Stanford,
CA 94305-4060, USA}
\altaffiltext{3}{BAER Institute, NASA Ames Research Center, Moffet Field,
CA 94043, USA}
\altaffiltext{4}{Department of Physics, New Jersey Institute of Technology,
Newark, NJ 07102, USA}

\begin{abstract}
Helioseismic and magnetohydrodynamic waves are abundant in and above
sunspots. Through cross-correlating oscillation signals in the photosphere
observed by the {\it SDO}/HMI, we reconstruct how waves propagate away 
from virtual wave sources located inside a sunspot. In addition to 
the usual helioseismic wave, a fast-moving wave is detected traveling 
along the sunspot's radial direction from the umbra to about 15~Mm beyond 
the sunspot boundary. The wave has a frequency range of $2.5 - 4.0$~mHz 
with a phase velocity of 45.3~\kms, substantially faster than the typical 
speeds of Alfv\'{e}n and magnetoacoustic waves in the photosphere. 
The observed phenomenon is consistent with a scenario of that a 
magnetoacoustic wave is excited at approximately 5~Mm beneath the sunspot, 
and its wavefront travels to and sweeps across the photosphere with a 
speed higher than the local magnetoacoustic speed. The fast-moving wave, 
if truly excited beneath the sunspot's surface, will help open a new 
window to study the internal structure and dynamics of sunspots. 
\end{abstract}


\section{Introduction}
Oscillations, helioseismic waves, and different types of magnetohydrodynamic
(MHD) waves are abundant in sunspots' atmosphere, both at and above the 
photospheric level \citep{bog05}. Studying these various waves can help us 
understand the physical conditions, such as temperature, density stratification,
and magnetic field structure, in the atmosphere and even beneath the 
photosphere of sunspots. Inside sunspots, 5-minute oscillations, 
stronger than oscillations of other periods, are believed mostly due 
to acoustic waves traveling into sunspots from the surrounding areas 
\citep{bec72, tho85}. In the chromosphere, 3-minute oscillations 
become dominant and are often associated with umbral flashes \citep[e.g.,]
[]{nag07}. Running penumbral waves (RPWs) are a prominent phenomenon 
observed in the sunspot chromosphere, and they display as concentric 
ripples propagating from the umbra-penumbra boundary outward with a typical 
speed of $10-25$ \kms\, \citep{gio72, zir72}, although speeds as low as 
$4 - 12$ \kms\, \citep{mad15} and as fast as $28 - 65$ \kms\, \citep{kol09} 
were also reported.  The nature of the RPWs has been long debated, and in 
recent years it was suggested that these waves might represent
magnetoacoustic waves that propagate upward from the photosphere along 
inclined magnetic field lines \citep{blo07, jes13, mad15}. Other 
chromospheric observations also revealed waves propagating 
only inside sunspot umbrae with a speed of $45-60$ \kms, which appeared 
to terminate near the umbra and penumbra boundaries \citep{kob04, lia11}. 
The RPWs and the umbral waves are thought to be independent phenomena. 
More recent observations from {\it IRIS} ({\it Interface Region Imaging 
Spectrograph}) found that sunspot oscillations in the chromosphere 
and the transition region displayed the behavior of shock waves \citep{tia14}.

How helioseismic waves, including both $f$-mode (surface gravity wave)
and $p$-mode (acoustic wave), interact with sunspots has been 
studied to better understand the sunspots' internal structure and dynamics, 
as well as wave properties in the presence of magnetic field. 
\citet{cam08, cam11} reconstructed from observations how the $f$-mode 
wave passes through a sunspot surface like a plane wave, and modeled 
numerically the sunspot's semi-empirical internal structures through 
comparing the simulated wave with the observationally reconstructed wave. 
Using a similar helioseismic technique to reconstruct the waves' 
propagation, \citet{zha11a} and \citet{yan12} measured acoustic wavefunctions 
scattered by sunspots, and \citet{cho12} studied the interference fringes 
of acoustic waves around sunspots. From a different perspective, 
\citet{zha11b} investigated how $f$- and $p$-mode waves, originating 
from virtual point sources, interacted with a sunspot before, during, 
and after their encounters. Later, through numerical simulations, 
\citet{par11} studied how helioseismic waves from a point source 
interacted with a sunspot in the hope to shed light on the sunspot's 
internal dynamics and magnetic field structure.

In this Letter, employing a procedure similar to reconstruct waves traveling
from virtual point sources \citep{zha11b}, we report on a detection of 
fast-moving waves at the photospheric level, propagating radially from 
the sunspot umbra through the penumbra to the outside. The observed properties 
of this wave are consistent with a fast magnetoacoustic wave excited beneath 
the sunspot's surface, the wavefront of which sweeps across the photosphere. 
Although the cause of this subsurface wave source is not clear at this time, 
this new phenomenon may help to open a new window to diagnose 
the sunspots' interior. We introduce the data and analysis 
procedure in Sec.~2, present the results in Sec.~3, and discuss the nature 
and possible cause of the wave in Sec.~4.

\section{Observation and Analysis} 
In this Letter, we present analysis of a sunspot located inside active
region NOAA~11312 as one example; however, it is worth pointing out that 
we have obtained essentially the same results for several other sunspots.
We believe that the results reported here are a common phenomenon existing 
in most sunspots.

\begin{figure}[!t]
\epsscale{0.9}
\plotone{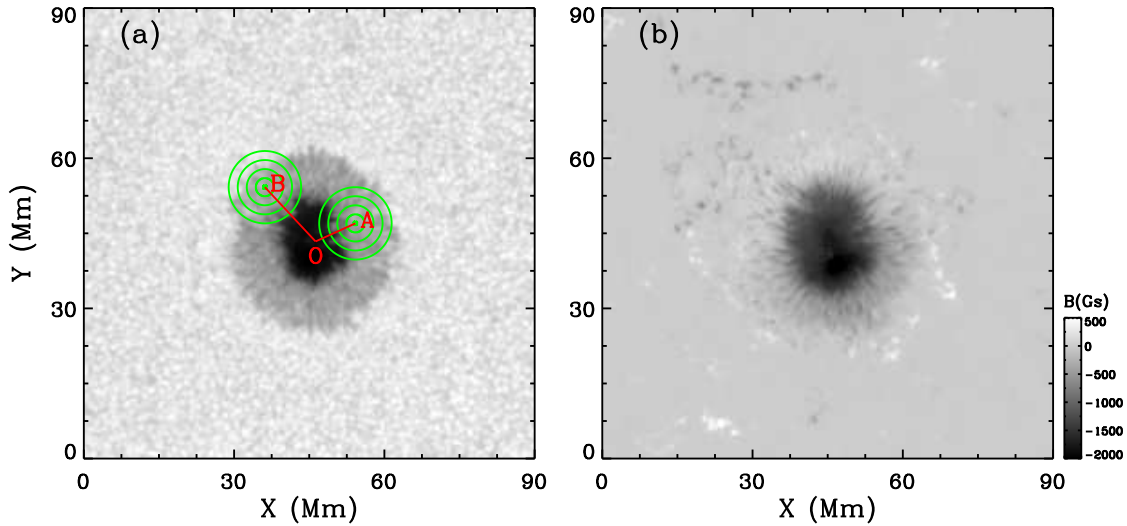}
\caption{(a) Continuum intensity and (b) magnetic field of the sunspot 
located in AR~11312. Panel (a) also shows a schematic plot illustrating
the data analysis scheme. Roughly, `$O$' is the geometric center of the 
sunspot, and lines `$OA$' and `$OB$' are along the sunspot's radial 
direction.}
\label{Ic_mag}
\end{figure}

As shown in Figure~\ref{Ic_mag}, this sunspot appeared to be round, 
which can be presumably considered as axisymmetric for helioseismic 
purposes, and stayed relatively stable during its disk passage 
from 2011 October 3 through October 17. In this study, we use the 
data taken by {\it SDO}/HMI \citep[{\it Solar Dynamics Observatory} / 
Helioseismic and Magnetic Imager,][]{sch12a, sch12b} between 00:00UT 
of October 8 and 23:59UT of October 12, a 5-day period during the sunspot's 
passage from about $30\degr$ east to $30\degr$ west of the central 
meridian. To prepare the data for analysis, we track the sunspot with 
its local rotation rate, and remap the data into heliographic coordinates 
using Postel's projection with the sunspot located at the center of 
the images. In order to assure that the observed phenomenon in this 
study is not due to artifacts or unknown calibration errors, we analyze 
four observables obtained by the {\it SDO}/HMI: Doppler velocity, continuum 
intensity, line-core intensity, and line-depth, as well as one observable 
from the {\it SDO}/AIA \citep[Atmospheric Imaging Assembly,][]{lem12}: 
1700\AA\, line intensity.  For all observables, we use running difference 
images of the original data without applying other signal filters so that 
the solar convection gets substantially suppressed while wave signals around 
the sunspot are not compromised.

For stochastic wavefields, cross-correlating oscillation signals observed 
at one location with those observed at other locations is a powerful tool 
to reconstruct how waves propagate away from this location to 
others. This method has been widely used in various research fields, 
including seismology \citep[e.g.,][]{sni10, der14} and helioseismology 
\citep[e.g.,][]{cam08, zha11b}.  The analysis procedure used in this study 
is the same as described by \citet{zha11b}, but the virtual wave sources 
are chosen inside the sunspot. The analysis scheme is illustrated in 
Figure~\ref{Ic_mag}a: all waves originating from a virtual source `$A$' 
can be reconstructed by cross-correlating the time series observed at 
`$A$' with the time series observed at all other locations inside the 
field of view. Waves originating from the virtual source `$B$' can be 
reconstructed similarly. However, practically, the waves constructed from 
a single source using a limited time period do not possess sufficiently high 
signal-to-noise ratio for further analysis. Because the sunspot is nearly 
axisymmetric, we can enhance the signal-to-noise ratio through averaging 
the wavefields obtained from sources `$A$' and `$B$' by rotating the wavefield 
from `$B$' to overlap directions of `$OA$' and `$OB$' and match locations 
`$A$' and `$B$'. The same azimuthal averaging process is repeated for 
waves reconstructed for all other locations inside the sunspot. 
Moreover, when we average the wavefields for final results, we separate 
the cases for virtual sources located inside the umbra and penumbra in 
case that waves from these two different entities have different 
properties. The resultant wavefields resemble the propagation of waves 
originating inside the sunspot umbra and penumbra. However, due to that 
the penumbra covers a large distance range, our results from the sunspot 
penumbra smear out the differences, if any, of waves initiated from the 
inner, middle, and outer penumbra.

\begin{figure}[!t]
\epsscale{1.0}
\plotone{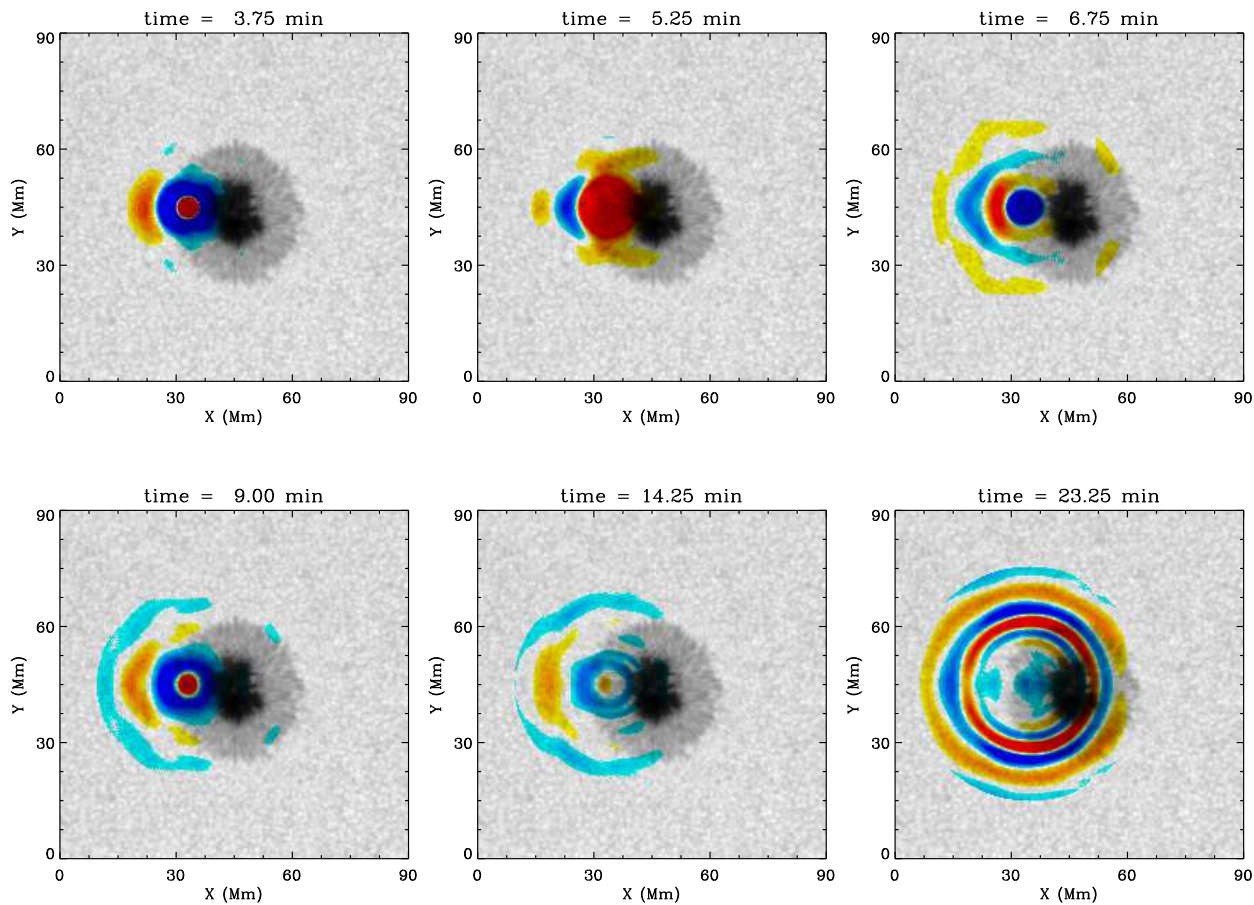}
\caption{Selected snapshots of the reconstructed waves propagating away 
from the sunspot penumbra, shown as foreground color images. The 
background black-and-white image shows continuum intensity of the 
studied region. The snapshots taken at 3.75, 5.25, 6.75, and 9.00 min 
show the fast-moving wave along the sunspot's radial direction, and the 
snapshots taken at 14.25 and 23.25 min mainly show the typical helioseismic 
waves expanding in all directions. An online movie is associated with
this figure.}
\label{waves}
\end{figure}

\section{Results}
\begin{figure}[!t]
\epsscale{0.95}
\plotone{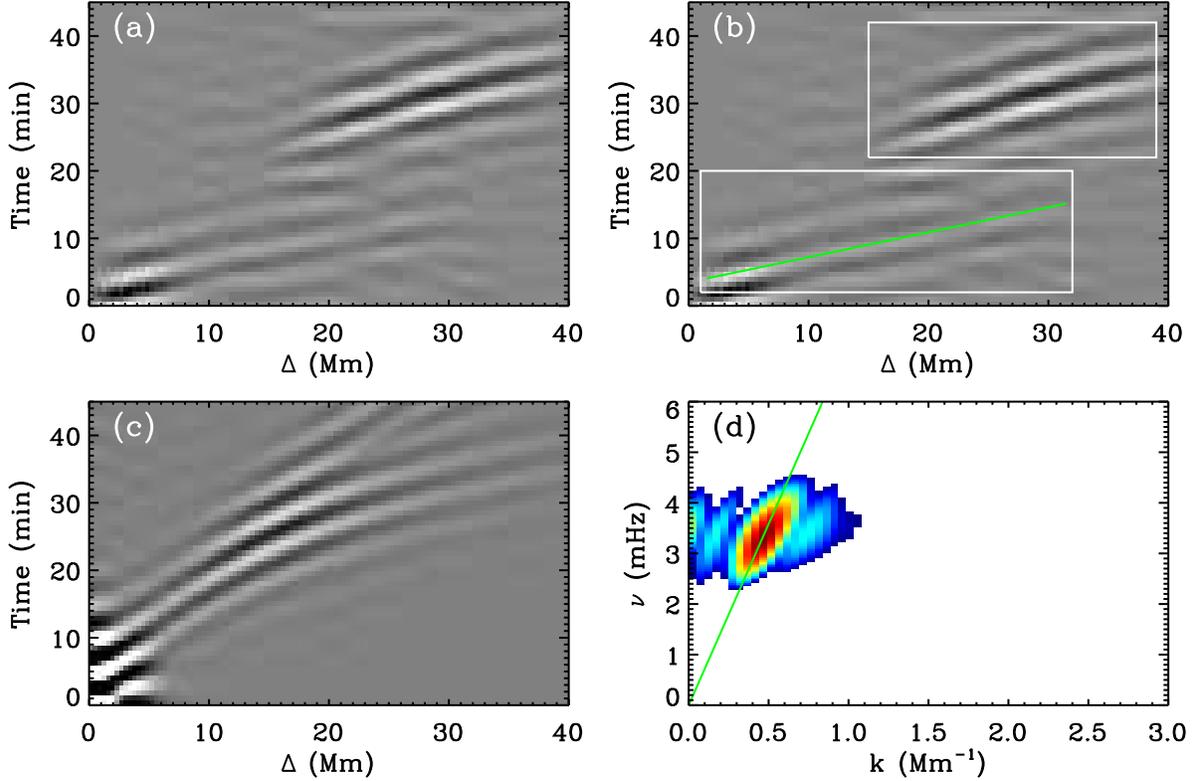}
\caption{(a) Time--distance diagram obtained along the sunspot's radially
outward direction. (b) Same as (a), but with two white boxes delimiting 
the areas of the fast-moving wave (lower box) and helioseismic waves (upper 
box). The green line shows a linear fitting of the fast-moving wave, 
corresponding to a speed of 45.3~\kms. (c) Time--distance diagram from 
a quiet-Sun region is shown for comparison. (d) Power-spectrum diagram 
for the fast-moving wave. The green line corresponds to the same phase 
speed as the line in panel (b). }
\label{td_pow}
\end{figure}

\begin{figure}[!ht]
\epsscale{0.85}
\plotone{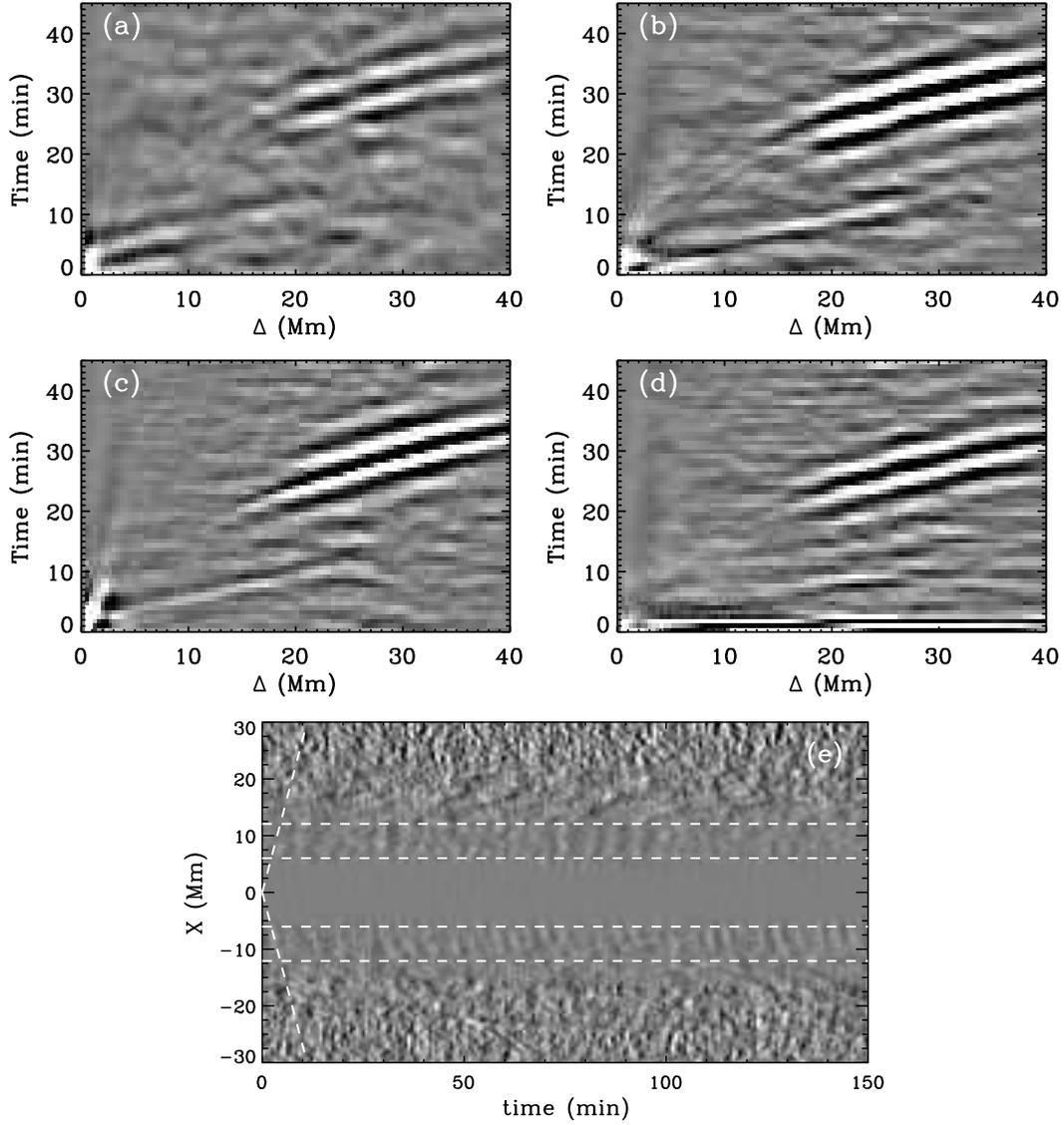}
\caption{Time--distance diagrams obtained in the same way as for the diagram 
shown in Figure~\ref{td_pow}a, but using data of (a) HMI continuum intensity,
(b) HMI line-core intensity, (c) HMI line depth, and (d) AIA 1700\AA. (e)
A time -- space plot obtained with a horizontal cut through the sunspot center, 
displayed in the HMI line-core intensity data. The inclined dashed lines 
indicate a propagation speed of 45.3~\kms. The waves of interest are confined 
between the white dashed lines, about $7 - 12$~Mm from the sunspot center.
Other wave patterns are visible beyond this range but are not related 
to the phenomenon discussed in the text. }
\label{all_tds}
\end{figure}

The results for the sunspot umbra do not differ significantly from the 
results for the penumbra regarding the properties of the wave 
but are much noisier. Therefore, in this Letter we 
focus on presenting and discussing only the results obtained 
when the virtual wave sources are located inside the penumbra. Also, the 
results obtained from the different HMI and AIA observables are qualitatively
similar, but the HMI Doppler velocity data give the best signal-to-noise 
ratio in our analysis. Figures~\ref{waves} and \ref{td_pow} show the 
results obtained using the Doppler velocity data, and Figure~\ref{all_tds} 
illustrates the results from the other observables.

Figure~\ref{waves} displays a few selected snapshots of the waveform obtained
for the virtual source located inside the penumbra, and the online movie  
shows the entire sequence. For the time lags from 3.75~min to 14.25~min, 
a fast-moving wave is seen propagating along the sunspot's radial direction 
from the penumbra to the outside of the sunspot. This wave is 
asymmetric with the strongest power and fastest traveling speed along 
the sunspot's radial outward direction, and weak or no power and slower 
speed for other directions. The wave signal terminates at approximately 
30~Mm away from the sunspot center, or about 15~Mm beyond the sunspot 
boundary. This fast-moving wave is clearly distinguishable from the usual 
helioseismic waves, which can be seen in the snapshots at 14.25~min and 
23.25~min. Actually, in our analysis, the helioseismic wave is first 
identified only at 12.00 min. 
%The helioseismic wave also exhibits a strong 
%asymmetry, faster toward the sunspot center and slower toward the outside. 

To better quantify the properties of the fast-moving wave, we obtain the 
time--distance relation of the wave by stacking together the cross-correlation
functions for different travel-time lags. Since the wave travels 
preferentially along the radially outward direction, only the time--distance 
diagram obtained along this direction is shown in Figure~\ref{td_pow}. 
There are two wave branches visible in the time--distance diagram 
(Figure~\ref{td_pow}a-b), with the lower branch corresponding to the 
fast-moving wave, and the upper branch corresponding to the usual helioseismic 
waves. By comparing this time--distance diagram with the one obtained from 
a quiet-Sun region following the same analysis procedure 
(Figure~\ref{td_pow}c), we find that the helioseismic waves from the 
penumbral sources are largely suppressed for short travel distances, 
and that the fast-moving wave is a phenomenon only associated with 
the sunspot.

The fast-moving wave appears to propagate with a nearly constant speed and 
shows little dispersion, i.e., its group velocity is similar to its phase 
velocity. We use a linear fitting to estimate its apparent phase velocity 
and get a speed of $45.3\pm1.7$ \kms, substantially faster than the speed 
of fast magnetoacoustic wave or Alfv\'{e}n wave, which is an order of 
10~\kms\, in the photosphere of sunspots. Taking the time--distance diagram 
of the fast-moving wave (the lower white box in Figure~\ref{td_pow}b), 
we calculate the power spectrum of this wave (Figure~\ref{td_pow}d). Due 
to the limited spatial and temporal scale of the wave, aliasing is hard 
to avoid when performing Fourier transform and calculating the power 
spectrum. The $k - \nu$ power-spectrum diagram shows a dominant power 
in the range of $2.5 - 4.0$~mHz, indicating that the oscillation 
frequency of this fast-moving wave falls into the category of 5-minute 
oscillation, same as the typical acoustic waves in the photosphere. The 
dominant power shows a phase velocity close to 45 \kms, in agreement with 
the fitting result in the space--time domain.

The fast-moving wave is also detected in the analysis results from the other 
HMI and AIA observables (Figure~\ref{all_tds}a-d), although the signal-to-noise
ratio varies among these observables. The properties of the observed
fast-moving wave are also slightly different in the characteristic
frequency and phase velocity, probably because different observables 
correspond to different atmospheric heights and the wave has different 
properties at different heights.

As a matter of fact, although the fast-moving wave is not apparent
in the HMI Doppler and continuum intensity data without applying the 
cross-correlation calculations, part of the fast-moving wave is directly 
visible in the HMI line-core and line-depth data. Figure~\ref{all_tds}e 
displays a time--space plot of the line-core data with a horizontal cut 
through the sunspot center, in which one can see herringbone patterns 
in the inner penumbra, i.e., about 7 -- 12~Mm away from the umbra center 
delimited in the dashed lines in Figure~\ref{all_tds}e. The herringbone 
patterns have a propagation speed of approximately 45~\kms, consistent 
with the estimate obtained from the cross-correlation calculations. 
Between 12 -- 20~Mm, the outer penumbra is dominated by slow-moving 
waves with an apparent speed of 6~\kms. These waves seem to be an extension
of some of the fast-moving waves, but they do not appear in the 
cross-correlation diagram (Figure~\ref{all_tds}b), implying these waves
do not share common sources with the fast-moving waves. Although not 
clearly visible in Figure~\ref{all_tds}e beyond 12~Mm, the fast-moving
waves can still be picked up by the cross-correlation analysis up to 
30~Mm from the sunspot center. What is the relation between the fast-moving
wave in the inner penumbra and the slow-moving wave in the outer
penumbra is worth further studying.

\section{Discussion}
Through cross-correlating stochastic oscillation signals observed at various 
locations inside and near a sunspot, we reconstruct waves propagating 
away from virtual point sources located inside the sunspot. It is found 
that after a virtual source is initiated and before typical helioseismic waves 
are visible, a fast-moving wave appears propagating outward along the 
sunspot's radial direction, with an apparent phase velocity of about 
45~\kms and a characteristic frequency between 2.5 and 4.0 mHz. This wave 
starts from the sunspot umbra and terminates at about 15~Mm beyond the 
sunspot boundary. In the inner penumbra, for about 1/4 of the whole 
traveling range, the wave is visible in the inner penumbra in the HMI 
line-core intensity data without applying cross-correlation calculations.
For the rest of the wave traveling range, and for HMI Doppler and 
continuum intensity data, the wave is only visible after cross-correlation 
analysis is applied. 

\begin{figure}[!t]
\epsscale{0.65}
\plotone{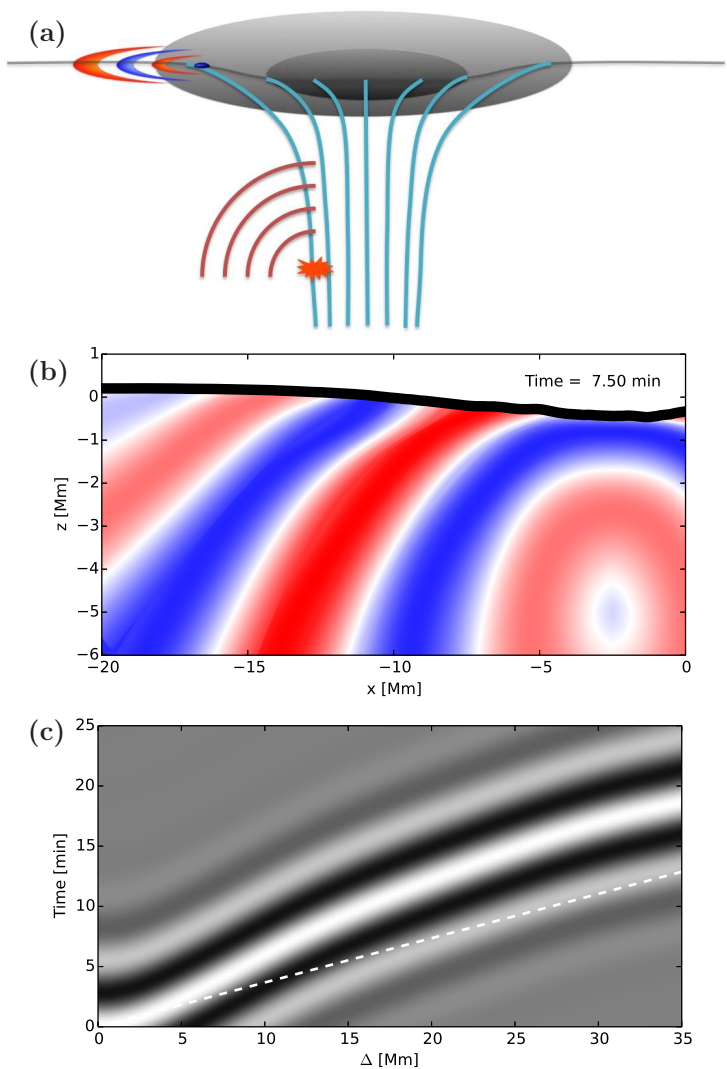}
\caption{(a) Schematic plot showing a scenario of that a subsurface disturbance
generates a helioseismic wave sweeping across the photosphere, forming 
the fast-moving wave observed in the photosphere. (b) Vertical snapshot 
showing wave propagating away from the disturbance, located at 2~Mm away 
from the sunspot's central axis at the depth of 5~Mm. The snapshot is taken 
at 7.5 min after the wave is excited. The wave shown in this figure is 
obtained from solving the ray-theory MHD equations using a realistic sunspot 
model for the frequency of 3~mHz. (c) Time--distance diagram seen in the 
photosphere for this wave. For comparison, a white dashed line, representing 
a speed of 45.3~\kms\, is plotted. }
\label{expect}
\end{figure}

What is the nature of this fast-moving wave, what causes it, and how to
explain its fast speed are interesting questions. Although 
Figure~\ref{waves} (and the online movie) seems to show the wave is 
a traveling wave on the surface, it is physically implausible to explain 
its propagation speed, which is substantially faster than the local Alfv\'{e}n 
wave and fast magnetoacoustic wave. The characteristic frequency of the wave 
implies that this wave falls into the category of acoustic waves, 
and the time--distance relation of the wave is a key factor to determine 
the nature of the wave. To explain this wave, we conjecture a disturbance 
occurring at approximately 5~Mm beneath the sunspot surface and 2~Mm
away from the sunspot's central axis (see Figure~\ref{expect}a). Acoustic 
waves are excited by this disturbance and expand toward all directions. 
Due to the stratified structure of the sound-speed profile as a function 
of depth, and the existence of the magnetic field that will alter, albeit 
slightly, the sound-speed profile both horizontally and vertically, 
the wavefront does not expand symmetrically toward all directions. The 
wavefront touches different locations of the photosphere at different times,
forming effectively a fast-expanding ellipse at the surface due to the 
asymmetric propagation speed. Numerous disturbance sources are located 
below the sunspot in a wide range of depths and distances to the central 
axis, and cross-correlation analysis eventually gives us a wave as 
seen in Figure~\ref{waves}. That is, it is likely that what we see at the 
surface are wavefronts of the waves, which are excited below 
sunspots, sweeping across the photosphere. Employing a realistic 
magnetohydrodynamic sunspot model \citep{rem09} and following the ray-path 
approximation with magnetic field \citep{mor08}, we calculate how a 
magnetoacoustic wave, at a frequency of 3.0~mHz, expand from the deep 
source and how its wavefront sweeps across the photosphere (see 
Figure~\ref{expect}). The time--distance relation from this calculation 
shows an approximately 40~\kms\, surface speed of the wavefront when it 
is 10~Mm horizontally away from the source on the surface. The difference in 
the apparent speeds between the model and observation may be due to the 
relatively weak magnetic field used in the model, or due to different 
depths of disturbance sources. We only present here our calculations of 
3.0 mHz, but we also recognize that the subsurface disturbances can excite 
waves with very broad frequency band. The frequency-dependent power
distribution at the subsurface locations is determined by the local
atmospheric properties, and is expected to be different from the power
distribution observed at the photosphere.

However, despite that this conjecture is promising in explaining the 
observations, there are clear difficulties. One difficulty is that it is 
unclear what causes those disturbances beneath sunspots? Some past 
helioseismic analyses indicated that the depth of 5~Mm was 
approximately where the sunspot's downdraft meets the upward flow
\citep[e.g.,][]{zha10}, and plasma collisions near this depth may cause 
disturbances that generate the magnetoacoustic waves. Other than that, 
currently there is no other evidence as we know indicating that this area 
may have plentiful wave sources. The other difficulty is, the 
magnetohydrodynamic processes between sunspots' subsurface and surface 
are rather complicated, and our equations prescribing the ray paths 
do not take into consideration of the subsurface flow field, the alteration 
of wave eignfunctions in the presence of strong magnetic field, and 
maybe other factors. The inclined magnetic field and fast flows in 
the penumbra further complicate the estimate of travel speed of the waves.  
Therefore, we do not expect that the picture given in Figure~\ref{expect} 
fully represents all details of the wave propagation through the sunspot 
region, but it is a useful first attempt to explain the observed phenomenon. 
Meanwhile, we cannot completely rule out other possible explanations, 
e.g., what we observe may be waves reflected back from the upper chromosphere 
\citep{cal13}.

It is likely that the chromospheric RPWs (see \S1) and the photospheric 
fast-moving waves have a common cause. When a disturbance occurs about 5~Mm 
below the sunspot surface near the sunspot's central axis, a magnetoacoustic 
wave sweeps across the photosphere in the sunspot and its immediate vicinity, 
and a fast-moving wavefront is detected. When the wave propagates along 
the inclined magnetic field from the subsurface to the photosphere and 
above, the RPW is observed in the chromosphere as the transverse projection 
of the magnetoacoustic wave \citep[e.g.,][]{blo07}. 

This newly detected fast-moving wave, although may not be a new type of 
wave, likely carries rich information from sunspots' subsurface area,
and may help to open a new window to diagnose sunspots' internal structure 
and dynamics.

\acknowledgments
{\it SDO} is a NASA mission, and HMI project is supported by NASA contract 
NAS5-02139 to Stanford University.

\end{document}